\documentclass{amsart}
\usepackage{amsmath}
\usepackage{amssymb}

\usepackage{graphicx}
\usepackage{color}

\newcommand{\matr}[1]{\mathbf{#1}}

\theoremstyle{plain}

\theoremstyle{definition}

\theoremstyle{remark}

\numberwithin{equation}{section}

\begin{document}

\title{A gauge theoretic approach to elasticity with microrotations}

\author{C. G. B\"OHMER}
\address{Department of Mathematics, University College London,\\
  Gower Street, London, WC1E 6BT, United Kingdom}
\email{c.boehmer@ucl.ac.uk}
\author{Yu.N. OBUKHOV}
\address{Department of Mathematics, University College London,\\
  Gower Street, London, WC1E 6BT, United Kingdom}
\email{obukhov@math.ucl.ac.uk}

\subjclass[2000]{(primary), (secondary)}

\begin{abstract}
We formulate elasticity theory with microrotations using the framework of gauge
theories, which has been developed and successfully applied in various areas of
gravitation and cosmology. Following this approach, we demonstrate the existence
of particle-like solutions. Mathematically this is due to the fact that our
equations of motion are of Sine-Gordon type and thus have soliton type
solutions. Similar to Skyrmions and Kinks in classical field theory, we can show
explicitly that these solutions have a topological origin. 
\end{abstract}

\maketitle

\section{Introduction}

\subsection{Elasticity and gauge theories}

In classical elasticity one assumes that the elements of a material continuum
are structureless points which are characterized by their positions only, 
whereas their internal properties (such as orientation, deformation etc)
are disregarded. 
This picture is generalized to the concept of continua with microstructure when 
a material body or fluid is formed of a large number of deformable particles 
whose (micro)elastic properties contribute to the macroscopic
dynamics of such a material medium. The corresponding mechanical model was 
developed very early by the Cosserat brothers \cite{coss} who replaced a 
structureless material point by a point that carries a triad of axes (an
oriented frame, in the modern geometrical language). 
This idea has led to a rich variety of models  which are known by
various different names like oriented or multipolar medium, asymmetric
elasticity, micropolar elasticity, see
e.g.~\cite{erik2,erik3,erik4,erin0,gr,mind,sch1,sch2,sch3,tou1,tou2,Eringen99}. 
This field is also related to other areas of continuum mechanics, like the 
theory of ferromagnetic materials, cracked media, liquid crystals 
\cite{degennes}, superfluids \cite{He3,Mermin,Shankar,Volovik} and
granular media (see, for example, application of the Cosserat model to 
the description of the grain size effects in metallic polycrystal \cite{Zeg}).

In general case, the 3-frame attached to material elements is deformable, thus
introducing nine new degrees of freedom in addition to the classical elasticity 
(3 rotations, 1 volume expansion, and 5 shear deformations); this is usually
called a micromorphic continuum \cite{Eringen99}. A special case is the
microstretch continuum for which the 3-frame can rotate and change volume 
(4 degrees of freedom: 3 microrotations and 1 microvolume expansion). Finally, 
in the micropolar continuum, the structured material point is rigid, thus
possessing 3 microrotation degrees of freedom. The last is the subject of the
Cosserat elasticity that considers media which can experience displacements
and microrotations. In turn, this theory has two important limiting cases, 
one with no microrotations which is the classical elasticity, and the second 
case where one assumes that the medium only experiences microrotations 
and no displacements. 
Models of the latter type have in fact a long history which can be traced back 
to MacCullagh in 1839, see~\cite{Whit,Unzicker}. Recently~\cite{re1,re2} such 
models have been investigated in the fully nonlinear setting and plane wave 
type solutions were explicitly constructed. The existence of such solutions 
is a highly non-trivial fact. In a linearised setting, similar solutions were 
investigated in~\cite{na1,na2,na3}.

Recent years have seen a revival of elasticity, in particular in the theory of
dislocations which has been analysed from a gauge theoretic point of view. The
gauge approach was developed and successfully applied in high energy physics
and in the theories of gravity \cite{fgfh,fhyo,mag}. The gauge 
approach in elasticity also has a long and rich history, see for example  
\cite{Ed1,Ed2,Mal1,Mal2,Kat1,Kat2,Kat3,Lazar09,LA,mlfh,lazar}. Of particular 
interest is the fact that the dislocations behave very much like particles.
{}From a more formal point of view, one of the open questions in this field 
of research is whether or not it is possible to find soliton type solutions 
which could justify the particle interpretation from a mathematical point of 
view. The main result of this paper is that we are able to find soliton type 
solutions in a particular model of elasticity motivated by gauge theories 
of gravity.

\subsection{Geometry and elasticity}

Our notation and conventions follow the book \cite{birk} where the modern 
differential geometry of manifolds is presented. Here we briefly summarize 
the notions and define objects needed for our current study. 

We consider an elastic medium which occupies the whole of $\mathbb{R}^3$, and 
we can identify material points with points in space. We use the Latin alphabet 
to label the coordinate indices, $i,j,k,\dots = 1,2,3$, say, we write $x^i$ for 
the spatial (holonomic) coordinates of $\mathcal{M}$, a 3-dimensional simply
connected manifold which we identify with the material points of the continuum.
This manifold occupies the entire 3-dimensional Euclidean space $\mathbb{R}^3$.
Greek indices refer to the (co)frame indices (also tangent space or anholonomic
indices), we use $\alpha,\beta,\dots = 1,2,3$ to label the frame covectors 
(1 forms) $\vartheta^{\alpha} = h^{\alpha}_i dx^i$ which we can refer to as the 
distortion in elasticity. We follow the Einstein summation convention whereby 
we sum over twice repeated indices. We denote the frame by $e_{\alpha} = 
h_{\alpha}^i\,\partial/\partial x^i$, such that $e_{\alpha}\rfloor\vartheta^{\beta}
= h_{\alpha}^i h^{\beta}_i = \delta_{\alpha}^{\beta}$. The $e_{\alpha}$ form the basis 
vectors of the tangent manifold $T\mathcal{M}$, and $\rfloor$ denotes the 
interior product. 

The Euclidean metric $g$ is defined on $\mathcal{M}$ as a scalar product $g: 
T\mathcal{M}\times T\mathcal{M} \rightarrow\mathbb{R}$. As usual, the values 
of this scalar product on the basis vectors introduce the metric tensor 
$g_{ij} = g(\partial_i, \partial_j)$, the components of which for the Cartesian 
coordinates read $g_{ij} = \delta_{ij} = {\rm diag}(+1,+1,+1)$. The components
of the metric $g_{\alpha\beta} = g(e_\alpha, e_\beta)$ in an arbitrary basis are 
obviously related to the frame components via $g_{\alpha\beta} = h_{\alpha}^i 
\delta_{ij} h_{\beta}^j$. One can think of $\vartheta^\alpha$ as a set of three 
mutually orthogonal and normal covectors which form a basis at any point 
$p \in T^{*}\mathcal{M}$. Each such basis covector has components 
$(\vartheta^{\alpha})_i$ and thus we can also view the object $h^{\alpha}_i$ 
as an orthogonal matrix. We can think of the metric of the deformed medium 
as the Cauchy-Green tensor in elasticity. Technically, the metric is used 
to raise and lower the indices, for example we obtain a covector-valued 
1-form from a vector-valued coframe as $\vartheta_\alpha = g_{\alpha\beta}
\,\vartheta^\beta$, etc. The components of the coframe comprise the $3\times 3$ 
matrix $h^{\alpha}_i$ which is inverse to matrix composed of the frame
components $h_{\alpha}^i$. From the definitions we directly obtain the following 
identity $h_i^{\alpha} = g^{\alpha\beta} \delta_{ij} h^j_{\beta}$.

One can draw a close analogy between gravity and elasticity, see the
instructive discussion \cite{fgfh}, for example. The force and the hyperforce 
(or the couple-force), as well as the stress (force per area) and the 
hyperstress (couple per area), caused by the force and hyperforce, respectively,
are the basic concepts in the continuum mechanics of structured media. The 
direct gravitational counterparts of the stress and hyperstress are the 
energy-momentum current and the hypermomentum (that includes spin, dilation 
and shear currents). The spacetime itself is geometrically modelled as a 
continuum deformed under the action of the sources with nontrivial ``stress'' 
(energy-momentum) and ``hyperstress'' (spin current, e.g.). Using this 
correspondence, it is possible to put both gravity theory and elasticity 
theory into the same differential-geometric framework in which the metric, 
coframe, and connection are the fundamental field variables. We refer the 
reader for more formalism and technical detail to the comprehensive review 
\cite{mag}.

When considering deformations of an elastic material, one places certain
compatibility conditions on the induced stress. These so called Saint-Venant
compatibility conditions are a form of integrability conditions so that the
stress can be expressed and the symmetric derivative of the displacement. In a
more geometrical language, by deforming the medium, we do not want to induce any
curvature into the material. This means that one assumes the Riemann curvature
tensor of the deformed medium to vanish identically. It is well known that these
two views are equivalent, see for instance~\cite{Am,So}. In the continuum 
theory of media with defects \cite{kroener2}, the elastic properties of a 
crystal are described by means of the non-Riemannian geometries in which
the notion of curvature is supplemented by a new geometrical quantity called 
torsion which was first introduced by Cartan in 1922. Therefore, we will 
analyse elastic materials using the language of the Riemann-Cartan geometries 
with torsion and we will stick to the well established requirement of having 
vanishing total curvature. The latter condition is sometimes called a second 
basic law in the differential geometry of media with defects \cite{kroener2}, 
and it has the physical meaning that the crystal structure is uniquely defined 
everywhere, see also \cite{Bilby,Kondo1,Kondo2}. In the following subsection 
we will collect the most important facts of such geometries.

\subsection{Geometries with vanishing Riemann curvature tensor}\label{geom}

Let us describe each material point of our elastic material by a coframe
$\vartheta^\alpha$ (a vector-valued 1-form) and a connection $\Gamma_\alpha
{}^\beta$ (a matrix-valued 1-form). Recall 
that the latter arises from the computation of the covariant derivative 
$\nabla_v$ along a vector field $v$ of the frame vectors, $\nabla_ve_\beta 
=\Gamma_\beta{}^\alpha(v)\,e_\alpha$, see Sec. C.1 of \cite{birk}. The connection 
is assumed to be metric-compatible which means the vanishing nonmetricity 
$Dg_{\alpha\beta} =0$, hence the connection 1-form is skew-symmetric, $\Gamma_{
\alpha\beta} = -\,\Gamma_{\beta\alpha}$ (where $\Gamma_{\alpha\beta} = \Gamma_\alpha
{}^\gamma\,g_{\beta\gamma}$). The coframe specifies the orientation of the 
orthonormal basis vectors at this point, while the connection determines 
how an arbitrary vector is parallelly transported near this point. The 
vanishing of the Riemann-Cartan curvature tensor (a matrix valued 2-form) 
takes the form 
\begin{align}
R_\alpha{}^\beta := d\Gamma_\alpha{}^\beta + \Gamma_\sigma{}^\beta\wedge 
\Gamma_\alpha{}^\sigma \equiv 0,\label{Rzero}
\end{align}
where $d$ denotes the exterior derivative and $\wedge$ is the exterior
(alternating) product. In geometries where the Riemann-Cartan curvature 
identically vanishes the connection is often referred to as a Weitzenb\"ock 
connection. The condition $R_\alpha{}^\beta \equiv 0$ is known in the literature 
by various names like teleparallel, fernparallel of distant parallel geometry. 
Geometrically this condition means that the notion of parallelism is
no longer a local statement, but can be defined globally, this means on the
entire manifold. A straight line connecting two material points defines an
absolute notion of parallelism. Physically the condition (\ref{Rzero}) has
the meaning that the structure of an elastic medium is uniquely defined 
everywhere in $\mathcal{M}$, \cite{Kondo1,Kondo2,kroener2,Bilby}. The only 
nonvanishing geometrical object in such geometries is Cartan's torsion 
(a vector valued 2-form) defined by
\begin{align}\label{tor}
  T^\alpha = d\vartheta^\alpha + \Gamma_\beta{}^\alpha\wedge\vartheta^\beta.
\end{align} 

Let $\varepsilon = \vartheta^1\wedge\vartheta^2\wedge\vartheta^3$ denote the 
volume 3-form, then the dual forms of the products of the coframe are defined by
\begin{align}\label{eta1}
  \varepsilon_\alpha &:= *\vartheta_\alpha =e_\alpha\rfloor\varepsilon,\\
  \varepsilon_{\alpha\beta} &:= *(\vartheta_\alpha\wedge\vartheta_\beta) =
e_\beta\rfloor\varepsilon_\alpha,\label{eta2}\\
  \varepsilon_{\alpha\beta\gamma} &:= *(\vartheta_\alpha\wedge\vartheta_\beta
\wedge \vartheta_\gamma) = e_\gamma\rfloor\varepsilon_{\alpha\beta}.\label{eta3}
\end{align}
The operator $*$, introduced by the formulas above, is called the Hodge 
operator.The dual forms satisfy the following useful identities
\begin{align}\label{veta1}
  \vartheta^\beta\wedge\varepsilon_\alpha &= \delta_\alpha^\beta\varepsilon,\\
  \vartheta^\beta\wedge\varepsilon_{\mu\nu} &= \delta^\beta_\nu\varepsilon_{\mu}
- \delta^\beta_\mu\varepsilon_{\nu},\label{veta2}\\   
  \vartheta^\beta\wedge\varepsilon_{\alpha\mu\nu}&= \delta^\beta_\alpha
\varepsilon_{\mu\nu} + \delta^\beta_\mu\varepsilon_{\nu\alpha} +
\delta^\beta_\nu\varepsilon_{\alpha\mu}.\label{veta3}
\end{align}
The object $\varepsilon_{\alpha\beta\gamma}$ is the totally antisymmetric
Levi-Civita symbol with $\varepsilon_{123}=1$.

\subsection{Elastic invariants}

Since $T^\alpha$ is the only nontrivial geometric object, it is therefore the
only non-trivial object characterising the deformations of the medium. It is
natural to decompose it into its irreducible pieces which will serve as our
building blocks of elastic invariants. The torsion $T^\alpha = {\frac 12}
T^\alpha{}_{\mu\nu}\,\vartheta^\mu\wedge\vartheta^\nu$ is represented as a rank
3 tensor which is skew-symmetric in one pair of indices, $T^\alpha{}_{\mu\nu}
= - T^\alpha{}_{\nu\mu}$. It has 9 independent components. Since we work in
$\mathbb{R}^3$, it is also natural to consider the Hodge dual of torsion which 
can be regarded as a $3\times 3$ matrix with no {\it a priori} symmetries. At any point we can view a $3\times 3$ matrix as an element of a 9 dimensional vector space $V$. Performing rotations which leave the metric invariant will change the components of this matrix, thus the group $\mathrm{SO}(3)$ acts on $V$ and we are interested in identifying invariant subspaces. Those give rise to the so-called irreducible pieces of either this matrix, or its Hodge dual, the torsion tensor. Therefore, let us define $\mathcal{T}^\alpha := *T^\alpha$.

The three irreducible pieces of torsion are defined by
\begin{align}
  {}^{(2)}\!T^\alpha &= {\frac 12}\,\vartheta^\alpha\wedge (e_\beta\rfloor
T^\beta), \label{T2}\\
  {}^{(3)}\!T^\alpha &= {\frac 13}\,*\!(\vartheta_\beta\wedge
T^\beta)\,\varepsilon^\alpha, \label{T3}\\
  {}^{(1)}\!T^\alpha &= T^\alpha -{}^{(2)}\!T^\alpha -
{}^{(3)}\!T^\alpha.\label{T1}
\end{align}
As we already explained, the metric is used for the raising and lowering of 
the indices, i.e., $\varepsilon^\alpha = g^{\alpha\beta}\varepsilon_\beta$ and
$\vartheta_\beta = g_{\beta\gamma}\vartheta^\gamma$. 

The irreducible pieces (\ref{T2})-(\ref{T1}) can be conveniently understood
in tensor language as the skew-symmetric part, the trace part, and the 
traceless symmetric part of the matrix, see the discussion below and in 
particular the explicit formulas (\ref{Ts3})-(\ref{Ts1}).

It is important to note that there is a gauge freedom in choosing the variables
$\{\vartheta^\alpha,\Gamma_\alpha{}^\beta\}$. All coframes are related to each
other by means of local rotations. Therefore, the same elastic medium can be
described equivalently by another pair
$\{\vartheta'^\alpha,\Gamma'_\alpha{}^\beta\}$ which is related to the former by
means of the gauge transformation
\begin{align}
  \vartheta'^\alpha &= L^\alpha{}_\sigma\,\vartheta^\sigma, \label{traV}\\
  \Gamma'_\beta{}^\alpha &= L^\alpha{}_\sigma\,\Gamma_\rho{}^\sigma\,
(L^{-1})^\rho{}_\beta + L^\alpha{}_\sigma\,d(L^{-1})^\sigma{}_\beta,\label{traG}
\end{align}
where $L^\alpha{}_\beta$ is an arbitrary $3\times 3$ rotation matrix, this means
$L \in \mathrm{SO}(3)$. The covariant condition $R_\alpha{}^\beta \equiv 0$
is preserved by this gauge transformations, whereas torsion $T^\alpha$
transforms covariantly, $T'^\alpha = L^\alpha{}_\sigma\,T^\sigma$. 

Making use of the local transformations (\ref{traV})-(\ref{traG}), we can 
eliminate either of the variables in the pair $\{\vartheta^\alpha,\Gamma_\alpha
{}^\beta\}$. The ability to do this is closely to our assumption of a globally 
flat manifold~(\ref{Rzero}). In particular, in~\cite{re1}, the teleparallel 
connection is gauged away $\Gamma_\alpha{}^\beta = 0$, and the coframe 
$\vartheta^\alpha$ is left as the only variable. Alternatively in~\cite{re2} 
the coframe is gauged to the standard constant values $h^\alpha_i = 
\delta^\alpha_i$ and the only variable is the connection, in which case 
once can write
\begin{align}\label{gam}
  \Gamma_\beta{}^\alpha =\Lambda^\alpha{}_\sigma\,d(\Lambda^{-1})^\sigma{}_\beta.
\end{align}
Here $\Lambda \in \mathrm{SO}(3)$ parametrises a teleparallel connection.
We will refer to this choice of variables as the connection gauge since the only
dynamical variable is the connection.

\section{Basic equations of the model}

\subsection{The potential energy}

The natural potential energy is based on the sum of the three quadratic
microrotational elastic invariants constructed from the nonvanishing torsion
\begin{align}
  V = \sum\limits_{i=1}^3\,c_i\,{}^{(i)}\!T^\alpha\wedge *T_\alpha,\label{V0}
\end{align}
where $c_i$, $i=1,2,3$, are non-negative constants which are physically interpreted 
as elastic moduli. The non-negativity of the material constants $c_i$ is a 
consequence of the condition of the positive semi-definiteness of the 
microrotational elastic energy, $V \ge 0$, in complete agreement with \cite{LA}.

In general, $V= V(T^\alpha)$ can depend arbitrarily on the torsion, but the 
choice of the quadratic function (\ref{V0}) yields the linear constitutive law, 
\cite{LA,mlfh,lazar}. When our medium occupies the whole of $\mathbb{R}^3$, 
the three quadratic terms in (\ref{V0}) are not independent because of the 
teleparallel condition (\ref{Rzero}). Recall that there is an identity 
(see Eq.~(5.9.18) of~\cite{mag}) that is valid in $n$-dimensional space:
\begin{align}
  \tilde{R}^{\alpha\beta}\wedge\varepsilon_{\alpha\beta} = 
  R^{\alpha\beta}\wedge\varepsilon_{\alpha\beta} - [{}^{(1)}\!T^\alpha
- (n-2){}^{(2)}\!T^\alpha - {\frac 12}\,{}^{(3)}\!T^\alpha]\wedge *T_\alpha
  - d\left(2\vartheta^\alpha\wedge *T_\alpha\right).\label{T2id}
\end{align}
The tilde denotes the Riemannian geometric objects, i.e. those constructed from
the Christoffel symbols of the corresponding Riemannian metric of the manifold.

Taking into account that the Riemann-Cartan curvature vanishes due to the
teleparallel constraint (\ref{Rzero}) and that by assumption the elastic
medium is embedded in a flat Euclidean space with $\tilde{R}^{\alpha\beta}=0$,
for $n=3$ the identity (\ref{T2id}) allows us to express the square of the first 
(tensor) irreducible part in terms of the trace and axial trace squares:
\begin{equation}
{}^{(1)}\!T^\alpha\wedge *T_\alpha = {}^{(2)}\!T^\alpha\wedge *T_\alpha 
+ {\frac 12}\,{}^{(3)}\!T^\alpha\wedge *T_\alpha
- d\left(2\vartheta^\alpha\wedge *T_\alpha\right).\label{TT}
\end{equation}

In tensor language, using (\ref{T2}), (\ref{T3}), and (\ref{veta1}),
(\ref{veta2}), we have
\begin{align}
{}^{(2)}\!T^\alpha\wedge *T_\alpha &= {\frac 12}\,T^\nu{}_{\alpha\nu}
T_\mu{}^{\alpha\mu}\,\varepsilon,\label{T2sq}\\
{}^{(3)}\!T^\alpha\wedge *T_\alpha &= {\frac 1{12}}\,(T_{\mu\rho\sigma}
\varepsilon^{\mu\rho\sigma})^2\,\varepsilon.\label{T3sq}
\end{align}

It is straightforward to establish relation between the irreducible parts
of the torsion 2-form $T^\alpha$ and its dual 1-form $\mathcal{T}^\alpha$.
We have $\mathcal{T}^\alpha = \mathcal{T}^{\alpha\beta}
\,\vartheta_\beta$, denote the matrix $\mathcal{T}^{\alpha\beta}$ by $\matr{T}$,
and we find explicitly the decomposition of the second
rank tensor 
\begin{align}
\mathcal{T}^{\alpha\beta} &= {}^{(1)}\!\mathcal{T}^{\alpha\beta}
+ {}^{(2)}\!\mathcal{T}^{\alpha\beta} + 
{}^{(3)}\!\mathcal{T}^{\alpha\beta}\label{Tstar},\\
{\matr{T}} &= {}^{(1)}\!{\matr{T}} 
+ {}^{(2)}\!{\matr{T}} + {}^{(3)}\!{\matr{T}},
\end{align}
into the trace, antisymmetric and traceless symmetric parts:
\begin{alignat}{2}
&{}^{(3)}\!\mathcal{T}^{\alpha\beta} = {\frac 16}\,T_{\mu\rho\sigma}
\varepsilon^{\mu\rho\sigma}\,g^{\alpha\beta}, &\qquad &{}^{(3)}\!\matr{T} 
= \frac{1}{3}\,(\rm{tr}\matr{T})\,\matr{I},\label{Ts3}\\
&{}^{(2)}\!\mathcal{T}^{\alpha\beta} = {\frac12}\,\varepsilon^{\alpha\beta\mu}
\,T^\nu{}_{\mu\nu}, &\qquad &{}^{(2)}\!\matr{T}{}^{\mathrm{T}} = 
-\,{}^{(2)}\!\matr{T}\label{Ts2}\\
&{}^{(1)}\!\mathcal{T}^{\alpha\beta} = {\frac 12}\,T^{(\alpha}{}_{\mu\nu}
\varepsilon^{\beta)\mu\nu}-{\frac 16}\,T_{\mu\rho\sigma}\varepsilon^{\mu\rho\sigma}
\,g^{\alpha\beta}, &\qquad &{}^{(1)}\!\matr{T}{}^{\mathrm{T}} =
{}^{(1)}\!\matr{T},\quad\rm{tr}\,{}^{(1)}\!{\matr{T}} = 0.\label{Ts1}
\end{alignat}

\subsection{The choice of the gauge}

In the recent papers \cite{re1,re2}, the propagation of waves was studied
in the structured medium under consideration. The results obtained describe 
the elastic waves in the two complementary pictures which arise due to the 
two different choices of the variables allowed by the gauge freedom 
(\ref{traV})-(\ref{traG}). In~\cite{re1}, the coframe gauge is used when 
the connection is zero (completely gauged out). The components of the coframe 
are the only dynamical variables. Geometrically, they represent an arbitrary 
orthogonal $3\times 3$ matrix. Since the connection is trivial in this gauge, 
the torsion 2-form reduces to $T^\alpha = d\vartheta^\alpha$. Using the 
components of the coframe and frame, $\vartheta^\alpha = h^\alpha_idx^ i$, 
$e_\alpha = h_\alpha^ i\partial_i$, we have for the holonomic torsion
\begin{equation}
T^i = h_\alpha^iT^\alpha =  h_\alpha^id(h^\alpha_j)\wedge dx^j =
h_\alpha^i\partial_kh^\alpha_j\,dx^k\wedge dx^j. \label{Ti}
\end{equation}
The dual 1-form then reads
\begin{equation}
\mathcal{T}^i = \mathcal{T}^i{}_j\,dx^j = h_\alpha^i\partial_k
h^\alpha_l\,\varepsilon^{kl}{}_jdx^j. \label{Tsi}
\end{equation}
Here $\varepsilon^{kl}{}_j = h^k_\alpha h^l_\beta h^\gamma_j\epsilon^{\alpha\beta}
{}_\gamma$ is the Levi-Civita tensor with respect to the holonomic coordinate 
basis. 

A complementary approach is developed in~\cite{re2}, where the only dynamical
variable is the connection, whereas the coframe is fixed to its trivial 
constant value $\vartheta^\alpha = \delta^\alpha_i\,dx^i$. The torsion
(\ref{tor}) reduces in this gauge to $T^\alpha = \Gamma_\beta{}^\alpha\wedge
\vartheta^\beta$, and since in the Weitzenb\"ock space the connection is 
given by (\ref{gam}), we have
\begin{equation}
T^i = \delta^i_\alpha\,T^\alpha =  \delta^i_\alpha\Lambda^\alpha{}_\sigma
\,d(\Lambda^{-1})^\sigma{}_\beta\,\delta^\beta_j\,dx^j.
\end{equation}
Denoting $u^i_\sigma := \delta^i_\alpha\Lambda^\alpha{}_\sigma$, we thus have
\begin{equation}
T^i =  u_\alpha^id(u^\alpha_j)\wedge dx^j =
u_\alpha^i\partial_ku^\alpha_j\,dx^k\wedge dx^j. \label{TiG}
\end{equation}
This is completely equivalent to (\ref{Ti}), with the difference that the 
orthogonal matrix $h^i_\alpha$, representing the coframe, is replaced with
the orthogonal matrix  $\matr{u} := u^i_\alpha$, representing the connection. 

In~\cite{re2}, there are two technical deviations as compared to the model
developed in~\cite{re1}. The first deviation is the different choice of the
variables mentioned above. Noticing that $(\matr{u}^{\mathrm{T}}\partial
\matr{u})^i{}_{kj} = u_\alpha^i\partial_ku^\alpha_j$ is skew-symmetric in $i,j$ 
(being an element of the Lie algebra of the orthogonal group), in~\cite{re2} 
one chooses as a basic variable a new $3 \times 3$ matrix
\begin{align} 
  \matr{A} = \star(\matr{u}^{\mathrm{T}}\partial\matr{u}),
\end{align}
which has no a priori symmetries. This $\star$ denotes the dualization operator
which relates skew symmetric matrices to vectors, similar to the Hodge $*$
operator. Writing out the indices explicitly, this matrix can be written in the
following way
\begin{equation}\label{Aij}
  A_{lk} = {\frac 12}\,\varepsilon_{li}{\,}^j\,u_\alpha^i\partial_ku^\alpha_j.
\end{equation}
This quantity is known in the polar elasticity theory under various different
names: the Nye tensor \cite{kroener}, the wryness tensor \cite{erik}, the 
second Cosserat deformation tensor \cite{Eringen99}, the third order right 
micropolar curvature tensor \cite{Neff}, or torsion-curvature tensor (note that
this is somewhat a misnomer since it neither directly relates to torsion nor 
curvature in the sense of differential geometry). The inverse of (\ref{Aij})
yields the components of the torsion tensor (\ref{TiG})
\begin{equation}
  u_\alpha^i\partial_ku^\alpha_j = A_{lk}\,\varepsilon^{li}{}_j.\label{Ainv}
\end{equation}
Substituting this into (\ref{TiG}), we can find a relation between the the 
second rank tensor $\matr{T}$ (which we can view as a matrix in $\mathbb{R}^3$) 
to the matrix $\matr{A}$, so that
\begin{equation}
  \matr{T} = \matr{A} - \rm{tr}(\matr{A})\,\mathbf{I}.\label{TA}
\end{equation}

The second deviation of~\cite{re2} from~\cite{re1} is a different structure
of the Lagrangian. Namely, the Lagrangian of a rotationally elastic medium
is assumed in~\cite{re2} to be a function of orthogonal matrix only.

\subsection{Kinetic energy and total Lagrangian}

The Lagrangian 4-form $V = L\,dt\wedge\varepsilon$ is constructed by taking 
$L = L(u,\partial_0u,\partial_iu)$ as a quadratic function of the irreducible
parts of $A_{lk}$. In addition, the kinetic term is chosen to be
\begin{equation}
L^{\rm kin} = {\frac 12}{\rm tr}(\dot{u}^{\rm{T}}\dot{u}) = {\frac 12}
(\partial_t u^i_\alpha)(\partial_t u_i^\alpha).\label{Lkin}
\end{equation}
This form of the kinetic energy is well motivated, since when linearised it
yields angular kinetic energy or rotation energy. Strictly speaking, in 
general the kinetic energy should be constructed by taking the microinertia
into account \cite{Eringen99,maugin}.

Introducing, analogously to (\ref{Aij}), the velocity of the deformations 
of the material continuum
\begin{equation}
A_{lt} = {\frac 12}\,\varepsilon_{li}{\,}^j\,u_\alpha^i\partial_t
u^\alpha_j,\label{A0k}
\end{equation}
and denoting $A^{lt} = g^{lk}A_{kt}$, we can recast (\ref{Lkin}) as
\begin{equation}
L^{\rm kin} = {\frac 12}(\partial_t u^i_\alpha)(\partial_t u_i^\alpha)
= A_{lt}A^{lt}.\label{Lkin2}
\end{equation}

Taking into account the identity (\ref{TT}), the potential energy contains
just two quadratic invariants. As a result, the general Lagrangian reads 
\begin{equation}\label{Ltot}
L= L^{\rm kin}(A_{lt}) - L^{\rm pot}(A_{lk}),\qquad L^{\rm pot}(A_{lk}) = 
{\lambda}_1\,(A^k{}_k)^2 + {\lambda}_2\,A_{[lk]}A^{[lk]}.
\end{equation}
The coupling constants $\lambda_1 = {\frac 43}(c_3 + {\frac 12}c_1)$ and 
$\lambda_2 = c_1 + c_2$ are constructed from the elastic moduli (\ref{V0}).

One can use different parametrizations of the orthogonal matrices. For 
example, in~\cite{re1}, the spinor parametrization was used. Here we find
it more convenient to describe an arbitrary orthogonal matrix with the 
help of the three real functions $\beta^i$ which is based on the 
identification of $SO(3)$ with the three-sphere. From these independent
constituents, an orthogonal matrix is constructed as follows
\begin{equation}
u_\gamma^i = \left(\delta_j^i + 2\beta_j\beta^i - 2\delta_j^i\beta^2
+ 2\alpha\beta^k\varepsilon^i{}_{jk}\right)\delta^j_\gamma.\label{u}
\end{equation}
Here $\beta^i = \delta^{ij}\beta_j$, $\alpha^2 + \beta^2 = 1$ (with $\beta^2 
= \delta_{ij}\beta^i\beta^j$).
The inverse matrix reads
\begin{equation}
u^\gamma_i = \left(\delta^j_i + 2\beta^j\beta_i - 2\delta^j_i\beta^2
- 2\alpha\beta^k\varepsilon^j{}_{ik}\right)\delta_j^\gamma.\label{uin}
\end{equation}
Substituting (\ref{u}) and (\ref{uin}) into (\ref{Aij}) and (\ref{A0k}), 
we find
\begin{eqnarray}
A_{lk} &=& 2\left(\varepsilon_{lij}\beta^i\partial_k\beta^j +
\beta_l\partial_k\alpha
- \alpha\partial_k\beta_l\right),\label{Abb1}\\
A_{lt} &=& 2\left(\varepsilon_{lij}\beta^i\partial_t\beta^j +
\beta_l\partial_t\alpha
- \alpha\partial_t\beta_l\right).\label{A0bb1}
\end{eqnarray}
For small $\beta$, the above formulas can be linearised, so that $A_{lk}
\approx -2\partial_k\beta_l$, $A_{lt} \approx -2\partial_t\beta_l$, and the
linearised model is described by the Lagrangian
\begin{equation}
L \approx 4\dot{\beta}^2 - 4{\lambda}_1({\rm div}\beta)^2 
- 2{\lambda}_2({\rm curl}\beta)^2.\label{Llin} 
\end{equation}
Here as usual ${\rm div}\beta = \partial_i\beta^i$, and $({\rm curl}\beta)_1
= \partial_2\beta_3 - \partial_3\beta_2$, etc. 

\subsection{The nonlinear equations of motion}

The complete nonlinear equations are more nontrivial. Denote the derivatives
\begin{equation}
H^{lk} = {\frac {\partial L^{\rm pot}}{\partial A_{lk}}},\qquad
H^{lt} = {\frac {\partial L^{\rm kin}}{\partial A_{lt}}}.\label{HH}
\end{equation}
Then the field equations read 
\begin{equation}\label{eqs}
(\partial_tH^{it} - \partial_kH^{ik})P_{ij} + 2(H^{it}Q_{tij} -
H^{ik}Q_{kij})=0,
\end{equation}
where we introduced 
\begin{eqnarray}
P_{ij} = \varepsilon_{ijl}\beta^l + {\frac 1\alpha}(\delta_{ij} -
\delta_{ij}\beta^2
+ \beta_i\beta_j),\label{Pij}\\
Q_{tij} = \left[\varepsilon_{ijl} - {\frac 1\alpha}(\delta_{ij}\beta_l 
- \delta_{il}\beta_j)\right]\partial_t\beta^l,\label{Q0}\\
Q_{kij} = \left[\varepsilon_{ijl} - {\frac 1\alpha}(\delta_{ij}\beta_l 
- \delta_{il}\beta_j)\right]\partial_k\beta^l.\label{Qk}
\end{eqnarray}

The equations (\ref{eqs}) are valid for {\it any} Lagrangian 
$L= L^{\rm kin}(A_{lt}) - L^{\rm pot}(A_{lk})$ with an arbitrary dependence on
the variables $A_{lt},A_{lk}$. However, for the specific choice of the 
quadratic Lagrangian (\ref{Lkin2}) and (\ref{Ltot}), we have explicitly
\begin{equation}
H^{it} = 2A^{it},\qquad H^{ik} = 2{\lambda}_1\,A^l{}_l\delta^{ik} + 
2{\lambda}_2\,A^{[ik]}.\label{HHq}
\end{equation}

The tensor (\ref{Pij}) is invertible and the inverse reads
\begin{equation}
(P^{-1})^{jk} = \alpha\,\delta^{jk} - \varepsilon^{jkn}\beta_n.\label{Pinv}
\end{equation}
One can easily check that $P_{ij}(P^{-1})^{jk} = \delta^k_i$. Multiplying 
(\ref{eqs}) with this inverse, we obtain another convenient form of the 
fields equations:
\begin{equation}\label{eqs2}
\partial_tH^{it} - \partial_kH^{ik} + 2(H^{jt}G_{tj}{}^i - H^{jk}G_{kj}{}^i)=0,
\end{equation}
where we have introduced
\begin{eqnarray}
G_{tj}{}^i &=& Q_{tjl}(P^{-1})^{li} =
\varepsilon_j{}^{il}\partial_t(\alpha\beta_l)
+ \beta^i\partial_t\beta_j - \beta_j\partial_t\beta^i,\label{G0}\\
G_{kj}{}^i &=&  Q_{kjl}(P^{-1})^{li} =
\varepsilon_j{}^{il}\partial_k(\alpha\beta_l)
+ \beta^i\partial_k\beta_j - \beta_j\partial_k\beta^i.\label{Gij}
\end{eqnarray}
Note that both objects are antisymmetric in $i,j$. 

The resulting system of nonlinear differential equations (obtained after
substituting (\ref{HHq}), (\ref{Pij})-(\ref{Qk}) and (\ref{Abb1}), (\ref{A0bb1})
into (\ref{eqs})) is quite nontrivial. It is possible, however, to find 
simple solutions under the additional assumptions. 

\section{Spherically-symmetric solutions -- the soliton}

Let us now look for the configurations with a 3-dimensional spherical symmetry. 
The corresponding ansatz reads
\begin{equation}
\label{sp} 
\beta^i = {\frac {x^i}r}\,\cos w,\qquad \alpha = \sin w,
\end{equation}
where the scalar function $ w =  w(t,r)$ depends on time and
on the radial variable $r = \sqrt{x_ix^i}$. 

Then (\ref{Abb1}) and (\ref{A0bb1}) yield
\begin{eqnarray}
A_{lk} &=& 2\left[\varepsilon_{lik}\,{\frac {x^i}{r^2}}\,\cos^2 w + 
{\frac {x_lx_k}{r^2}}\, w' - \sin w\cos w\left({\frac 
{\delta_{lk}}{r}} - {\frac {x_lx_k}{r^3}}\,\right)\right],\label{Asp}\\ 
A_{lt} &=& 2{\frac {x_l}r}\,\dot{ w}.\label{A0sp}
\end{eqnarray}
Hereafter the dot and the prime denote the derivatives with respect to time
and radius, respectively.

Accordingly, the trace and the skew-symmetric parts of (\ref{Asp}) are
\begin{eqnarray}
A^k{}_k &=& 2 w' - 4\,{\frac {\sin w\cos w}{r}},\label{At}\\
A_{[lk]} &=& -\,2\varepsilon_{lkn}\,{\frac {x^n}{r^2}}\,\cos^2 w.\label{Aa}
\end{eqnarray}
As a result, we find
\begin{eqnarray}
H^{it} &=& {\frac {4x^i}r}\,\dot{ w},\\ 
H^{ik} &=& 2{\lambda}_1
\,\delta^{ik}\left(2 w' - 4\,{\frac {\sin w\cos w}{r}}\right) 
- 4{\lambda}_2\,\varepsilon^{ikn}\,{\frac {x_n}{r^2}}\,\cos^2 w. 
\end{eqnarray}
It is straightforward to see that
\begin{eqnarray}
\partial_tH^{it} &=& {\frac {4x^i}{r}}\,\ddot{ w},\qquad 
H^{jt}G_{tj}{}^i = 0,\label{dHG} \\ 
G_{kj}{}^i &=& \varepsilon_{kj}{}^i\,{\frac {\sin w\cos w}{r}} +
\varepsilon_j{}^{il}\,{\frac {x_lx_k}{r^2}}\left({\frac {\sin w\cos w}{r}}
\right)'\nonumber\\
&& +\,(x^i\delta_{kj} - x_j\delta^i_k){\frac {\cos^2 w}{r^2}}.\label{Gsp}
\end{eqnarray}
Using these results, we find that the field equations (\ref{eqs2}) reduce to
\begin{equation}
{\frac {4x^i}{r}}\left[\ddot{ w} - {\lambda}_1\left( w'' +
{\frac 2r}\, w'\right) - {\frac 1{r^2}}\,U( w)\right] = 0,
\label{eqs3}
\end{equation}
where 
\begin{equation}
U( w) = \sin(2 w)[({\lambda}_2 - {\lambda}_1) +
({\lambda}_2 - 2{\lambda}_1)\cos(2 w)]. \label{U}
\end{equation}
Since the two terms in the round brackets are the Laplacian in spherical 
coordinates, this equation of motion can also be written in the neat form
$\ddot{w} - \lambda_1 \Delta w = U(w)/r^2$. Let us introduce a new function
$\varphi := w r$, and next let us rescale $r \mapsto \sqrt{\lambda_1}t$,
$\varphi \mapsto \sqrt{\lambda_1}\varphi$, then~(\ref{eqs3}) becomes
\begin{equation}
\ddot{\varphi}-\varphi'' + \frac{U(\varphi/r)}{\lambda_1^{3/2}\,r} = 0.
\label{eqs4}
\end{equation}
The resulting equation is closely related to the spherical `sine-Gordon'
equation, see for instance~\cite{De,Ol} and references therein. It is 
worthwhile to mention that the model of Kontorova-Frenkel \cite{KT1,KT2,KT3}
was the first theory discussing media with dislocations where the `sine-Gordon'
equation and its soliton solutions naturally appeared. The main
difference between previously studied spherical sine-Gordon equations and our
equation~(\ref{eqs4}) is that the nonlinearity carries an extra factor of $1/r$.
This is similar to angular momentum when studying the spherical Schr\"odinger
equation or Newton's equation. This additional factor has some interesting
implications. At large distances from the centre this term becomes negligible
and asymptotically we recover the wave equations. This analysis renders a
strong support for the existence of soliton solutions in our model with a 
localised configuration near the centre. Below we confirm this by numerical
integration. 

\begin{center}
\begin{figure}[!ht]
\includegraphics[width=.7\textwidth,height=0.3\textheight]{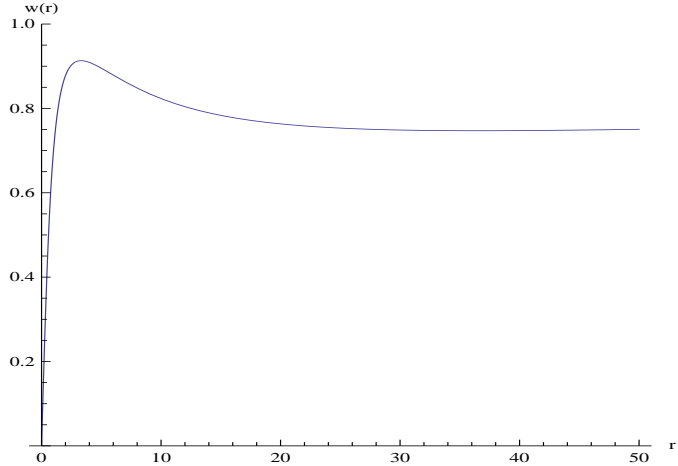}
\caption{Soliton solution of the rotational elasticity.}\label{fig1}
\end{figure}
\end{center}

In the static case, the equations of motion reduce to the single ODE
\begin{equation}
{\lambda}_1(r^2 w')' + U( w) = 0.\label{static}
\end{equation} 
For ${\lambda}_2 = 2{\lambda}_1$ this further reduces to $(r^2 w')'
+ \sin(2 w) = 0$, whereas when ${\lambda}_2 = {\lambda}_1$ 
one is left with $(r^2 w')' - {\frac 12}\sin(4 w) = 0$. The qualitative 
analysis of the equation (\ref{static}) reveals  the existence of static
solutions that vanish at $r =0$ and approach asymptotically $\pi/4$ at 
infinity for $r\rightarrow\infty$. 

Numerical integration is straightforward. The form of the solution depends
on the coupling constants and on the initial value of $w'(0)$. However, the
qualitative behaviour remains the same. As a specific example, Fig.~\ref{fig1}
presents the soliton for ${\lambda}_2 = {\lambda}_1$ when $w'(0) =1$.

In the static case we can introduce the new function $f$, defined by
\begin{equation}
w=\frac{1}{2}\arctan\{\sinh(f(\log(r)))\}
\end{equation}
which transforms~(\ref{eqs3}) into an autonomous second order differential
equation
\begin{equation}
 f_{\beta\beta} + f_{\beta}(1-\tanh(f)f_\beta) - 2\sinh(f) -4\tanh(f) +
2\frac{\lambda_2}{\lambda_1}(\sinh(f)+\tanh(f)) = 0,
\end{equation}
where $\beta=\log(r)$. This equation can now be analysed using standard
techniques from ordinary differential equations or dynamical systems, and we
find that this static system has three equilibrium points
\begin{equation}
f=0,\qquad \sinh f =
\pm\frac{\sqrt{\lambda_1}\sqrt{3\lambda_1-2\lambda_2}}{\lambda_1-\lambda_2},
\end{equation}
with eigenvalues $(0,-1)$ in all three cases which in turn yields interesting
(in)stability properties.

\section{Discussion}

The rotational elasticity model has many features similar to the model 
suggested by Skyrme~\cite{skyrme}. Although the Lagrangians are different,
the dynamics looks qualitatively the same. In particular, the remarkable
feature of the rotational elasticity is the existence of solitons which
are close relatives to the Skyrmions. The topological nature of such 
configurations (which we demonstrate below) guarantees the stability of 
solutions obtained in the sense that the variations of the coframe and
connection cannot change the value of the conserved charge. Moreover, the
moduli space of these solutions is obviously compact. 

Recently it was noticed~\cite{rand} that in the framework of the gauge approach 
in gravity and elasticity (that underlies the model under consideration, see 
Sec.~\ref{geom}) one can define an identically conserved current 3-form that 
gives rise to the topological charge that naturally classifies the field
configurations. Such a 3-current ``lives" on the 4-dimensional spacetime
that is constructed as a foliation where the $t=const$ slices coincide 
with the manifold $\mathcal{M}$. Specialising to the case of the Weitzenb\"ock 
geometry with the flat curvature (\ref{Rzero}), this topological current 
reduces to 
\begin{equation}\label{J}
J^{\rm top} = K_\alpha{}^\beta\wedge \tilde{D}K_\beta{}^\alpha + {\frac 23}
\,K_\alpha{}^\beta\wedge K_\beta{}^\gamma\wedge K_\gamma{}^\alpha.
\end{equation}
Here $K_\alpha{}^\beta = \Gamma_\alpha{}^\beta - \tilde{\Gamma}_\alpha{}^\beta$
is the contortion defined as the difference between the Riemann-Cartan and
the Riemannian connections. This 3-form is identically conserved on the 
4-spacetime, $dJ^{\rm top}\equiv 0$ in view of 
(\ref{Rzero}). As a result, we can construct the topological charge
\begin{equation}\label{Q}
{\mathcal{Q}} = {\frac {1}{96\pi^2}}\int K_\alpha{}^\beta\wedge
K_\beta{}^\gamma\wedge K_\gamma{}^\alpha. 
\end{equation}
The integral is taken over the 3-manifold $\mathcal{M}$. 
The topological nature of the charge (\ref{Q}) and current (\ref{J}) is 
manifested in the fact that the corresponding conservation law does not
depend on any dynamics of the fundamental fields $\vartheta^\alpha$ and
$\Gamma_\alpha{}^\beta$ \cite{rand}.

The integral (\ref{Q}) is taken over the whole 3-space and the result is a 
constant for configurations that we studied in the previous section when $w$ 
vanishes at the origin and approaches constant value at infinity. Specializing 
to the connection gauge (\ref{gam}), we have $\tilde{\Gamma}_\alpha{}^\beta =0$ 
(the coframe is ``gauged away'': $h^\alpha_i = \delta^\alpha_i$) and hence
\begin{eqnarray}
{\mathcal{Q}} &=& {\frac {1}{96\pi^2}}\int \Gamma_\alpha{}^\beta\wedge
\Gamma_\beta{}^\gamma\wedge \Gamma_\gamma{}^\alpha\nonumber\\ \label{Q1}
&=&  {\frac {1}{96\pi^2}}\int (\Lambda d\Lambda^{-1})_\alpha{}^\beta\wedge (
\Lambda d\Lambda^{-1})_\beta{}^\gamma\wedge(\Lambda d\Lambda^{-1})_\gamma{}^\alpha.
\end{eqnarray}
Fixing the asymptotic boundary conditions so that $\Lambda^\alpha{}_\beta =
\delta^\alpha_iu^i_\beta$ approaches unity near spatial infinity allows for
the standard compactification \cite{rand} of the space $\mathcal{M}\approx 
\mathbb{S}^3$. We thus verify that the charge (\ref{Q}), (\ref{Q1}) is a
degree (``winding number") of the maps $\Lambda: \mathbb{S}^3\rightarrow 
SO(3)$ which are classified by the integer numbers that are the elements 
of the third homotopy group $\mathbb{Z} = \pi_3(SO(3))$.
By direct computation we can verify that $\mathcal{Q} = 1$ for the soliton 
solution described above. 

Construction of a multi-soliton generalisation is an interesting problem. 
One can approach this along the lines proposed in \cite{rand}. As a first 
step, we can replace the ansatz (\ref{sp}) by taking 
instead of (\ref{u}) the orthogonal matrix that is a product of $N$ factors
\begin{equation}\label{u2}
u_\alpha^i = \left(u_{(1)}\cdot u_{(2)}\cdot\dots\cdot u_{(N)}\right)_\alpha^i.
\end{equation}
The dot denotes the usual matrix product. Such an ansatz would obviously 
generate a topological solution with the higher charge $\mathcal{Q} = N$. The 
construction of the multi-soliton configurations is a more nontrivial problem
which one could try to solve by using the instanton technique.   

The realization of the current theoretical model in the condensed matter 
systems is an interesting physical problem. The geometric constraint 
(\ref{Rzero}) of the vanishing curvature rules out the disclinations. It 
was shown in \cite{rand} that dislocations are also not relevant to the 
point-like soliton solutions. The solitons of this type are mentioned by
Unzicker \cite{Unzicker}, who calls them a ``Shankar monopole" \cite{Shankar}
that describes a point-like defect in a $A$-phase superfluid Helium-3,
\cite{He3,Mermin,Volovik}. As we mentioned already, our microrotational 
elasticity model belongs to the class of the so-called micropolar elasticity
theories. The relevant physical continua describe, in particular, liquid 
crystals with rigid molecules, superfluids, rigid suspensions, blood fluid 
with rigid cells, magnetic fluids, dust fluids etc. Eringen \cite{Eringen99} 
(see also references therein) extensively studied micromorphic, microstretch 
and micropolar elasticity models with other defects beyond the dislocations and 
disclinations. We refer again to the paper of Randono and Hughes \cite{rand}
where the point-like topological solution (called there a torsional monopole)
is depicted and a possible physical manifestations of such structures are
qualitatively related to certain nontrivial electronic behaviour of condensed 
matter systems with strong spin-orbit coupling such as (1+3)-dimensional
topological insulators. A careful quantitative analysis of such possibilities
will be left for a future study. 

\subsection*{Acknowledgments}

We thank Dmitri Vassiliev and Friedrich Hehl for discussions and advice. 
We also thank the referees for their valuable reports.

\end{document}